\documentclass[twocolumn,prb,floatfix,showpacs,citeautoscript]{revtex4}
\usepackage{graphicx}
\usepackage{amssymb}
\usepackage{pifont}

\begin{document}

%%%% Title of the paper %%%%%%%%%%
\title{Imaging of lateral spin valves with soft X-ray microscopy}

%%%% Authors from 1 institute %%%%
\author{O.~Mosendz}
\email{mosendz@anl.gov}
\affiliation{Materials Science Division, Argonne National Laboratory, Argonne, Illinois 60439, USA}

\author{G.~Mihajlovi\'c}
\affiliation{Materials Science Division, Argonne National Laboratory, Argonne, Illinois 60439, USA}

\author{J.~E.~Pearson}
\affiliation{Materials Science Division, Argonne National Laboratory, Argonne, Illinois 60439, USA}

\author{P.~Fischer}
\affiliation{Center for X-ray Optics, Lawrence Berkeley National Laboratory, Berkeley, California 94720, USA}

\author{M.-Y.~Im}
\affiliation{Center for X-ray Optics, Lawrence Berkeley National Laboratory, Berkeley, California 94720, USA}

\author{S.~D.~Bader}
\affiliation{Materials Science Division and Center for Nanoscale Materials, Argonne National Laboratory, Argonne, Illinois 60439, USA}

\author{A.~Hoffmann}
\email{hoffmann@anl.gov} \affiliation{Materials Science Division and
Center for Nanoscale Materials, Argonne National Laboratory,
Argonne, Illinois 60439, USA}

\date{\today}

%%%% Abstract %%%%%

\begin{abstract}
We investigated Co/Cu lateral spin valves by means of
high-resolution transmission soft x-ray microscopy with magnetic
contrast that utilizes x-ray magnetic circular dichroism (XMCD). No
magnetic XMCD contrast was observed at the Cu L$_3$ absorption edge,
which should directly image the spin accumulation in Cu. Although
electrical transport measurements in a non-local geometry clearly
detected the spin accumulation in Cu, which remained unchanged
during illumination with circular polarized x-rays at the Co and Cu
L$_3$ absorption edges.
\end{abstract}


%%%% Create FRONT MATTER %%%%%%%%%

\maketitle

%%%%%%%%%%%%%%%%%%%%%%%%%%%%%%%%%%

\section{Introduction}
Recent developments in spintronics showed that the use of spin
polarized currents can have a profound impact on applications such
as magnetic information processing and storage.  Furthermore, the
study of pure spin currents instead of spin polarized charge
currents is providing a promising new direction to advance the
present technology~\cite{Chappert-NP2008,Hoffmann-PSSC2007}. Pure
spin currents can be generated via a spin polarized charge current
injection from ferromagnets \cite{Johnson-PRL1985}, spin Hall
effects \cite{Kimura-PRL2007}, or spin pumping
\cite{Heinrich-PRL2003,Woltersdorf-PRL2007,Mosendz-PRB09}.  In order
to utilize pure spin currents it is necessary to understand their
propagation properties. To this end, the direct imaging of a
non-equilibrium spin accumulation that accompanies pure spin
currents can generate new insights into the dynamic behavior of
spins.  This became evident from prior research on semiconducting
systems, where such imaging is made possible due to strong
magneto-optic effects
~\cite{Awschalom-PRL98,Awschalom-SC04,Awschalom-NP05,Crooker-Science2005,Stephens-PRL2005}.
From such magneto-optical measurements it was discovered in
semiconductors that spin coherence times are relatively long and
injected spins can be transported with charge currents over
macroscopic distances;\cite{Awschalom-PRL98} spin Hall effects are
detectible;\cite{Awschalom-SC04,Awschalom-NP05} the imaging revealed
the flow-pattern of electrically injected spins including a spin
back-diffusion against the charge current at the draining
contact;\cite{Crooker-Science2005} and spin polarization can be
generated via spin-dependent reflection from metal-semiconductor
interfaces \cite{Stephens-PRL2005}.

For metallic spin transport systems such direct imaging of the spin
accumulation and spin currents remains elusive.  One reason is that
due to the smaller spin diffusion lengths, which are typically far
in the submicron range \cite{bass-review}, standard optical
techniques do not provide sufficient spatial resolution.
Nevertheless, it can be expected that successful imaging of the spin
accumulation in metals will greatly enhance our understanding of
spin currents and dynamics.  Some questions that could be
immediately addressed are how spin currents couple to charge drift
currents in metals \cite{Urech-NL2006} and what role inhomogeneous
spin injection plays for contacts whose sizes are comparable to the
spin diffusion length \cite{Hamrle-PRB2005}.

Soft x-ray microscopy with state-of-the-art Fresnel zone plates used
as high resolution optics~\cite{Fischer-IEEE08} is a promising
approach towards the goal of imaging spin accumulations in metals.
The spatial resolution, down to 10~nm, is sufficiently high and it
is possible to get magnetic contrast via x-ray magnetic circular
dichroism (XMCD) \cite{Chao-Nature2005,Fischer-MT2006}. Here we show
investigations of Co/Cu lateral spin valves, which were imaged using
magnetic transmission soft x-ray microscopy (MTXM) with circular
polarized x-rays for detecting XMCD from the spin accumulation in
Cu. Although the presence of a spin accumulation was verified with
non-local transport measurements, there was no detectable XMCD
contrast at the Cu L$_3$ absorption edge.

\section{Experimental}

The Co/Cu lateral spin valves were fabricated by means of e-beam
lithography utilizing a double-layer PMMA/PMGI resist on a SiN/Si
substrate. Selectively removing the Si enabled the SiN layer to be
used as a free-standing x-ray transparent membrane, which is
required to perform the MTXM imaging experiment. In order to
minimize detrimental absorption we used a 100-nm thick SiN layer.
Ohmic junctions for the lateral spin valves were formed by employing
a lithographically controlled e-beam resist undercut technique with
subsequent shadow mask evaporation of 25-nm Co and 80-nm Cu
\cite{Cord-JVST2006}. After their fabrication, the lateral spin
valves were covered by a 100~nm thick SiN protective layer deposited
via \emph{rf}-sputtering. Finally, the membrane windows were defined
with photolithography on the substrate back-side and the Si was
removed via wet etching in a 30~\% KOH solution.  This process
resulted in Co/Cu lateral spin valves sandwiched on both sides by
100~nm thick SiN layers forming a free-standing membrane.

All MTXM imaging was performed at beamline 6.1.2 (XM-1) of the
Advanced Light Source.  This microscope achieves 15~nm spatial
resolution using state-of-the-art Fresnel zone plates and enables
magnetic contrast via XMCD~\cite{Chao-Nature2005,Fischer-MT2006}.
During the measurements the synchrotron operated in a multibunch,
top-off mode delivering a continuous electron beam-current of
400~mA.  The beam position stability and the constant beam current
in the top-off mode was beneficial for the long-term stability of
the experiment, which accumulated x-ray images for two consecutive
days. For the MTXM imaging we chose incident energies of 777~eV and
932~eV for the Co and Cu L$_3$ absorption edges, respectively.
Images were obtained with circular polarized x-rays in remanence
after applying either positive or negative saturating magnetic
fields.  In order to obtain sufficient signal-to-noise ratios, we
accumulated 400 images per magnetization direction, which were added
together after aligning all images with rigid translations, in order
to correct for a slight drift of the microscope during the image
accumulation.

At XM-1 the samples are studied under ambient conditions, which
enabled us to readily make electrical connections to each of the
four lateral spin valves on every SiN membrane.  For samples with
small Co-contact separation we performed electrical transport
measurements using lock-in techniques and a maximum electrical
current of 0.5~mA. These transport measurements were performed with
the sample mounted into the microscope in order to allow for
simultaneous x-ray illumination and to study whether the impinging
x-rays have an effect on the transport measurements.

\section{Results and Discussion}
\begin{figure}
\includegraphics[width=8.6cm]{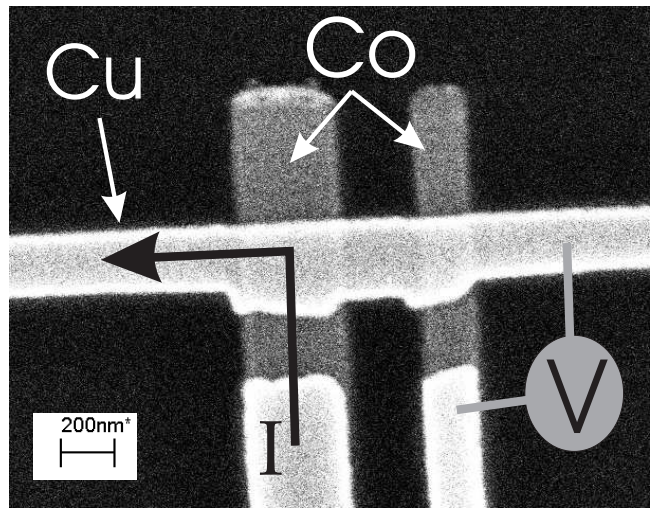}
 \caption{SEM image of a Co/Cu spin valve with 200-nm separation
between ferromagnetic electrodes.  The bright parts of the device
are the Cu wires, while the darker parts are the Co electrodes.}
\label{SEM}
\end{figure}

\begin{figure}
\includegraphics[width=8.6cm]{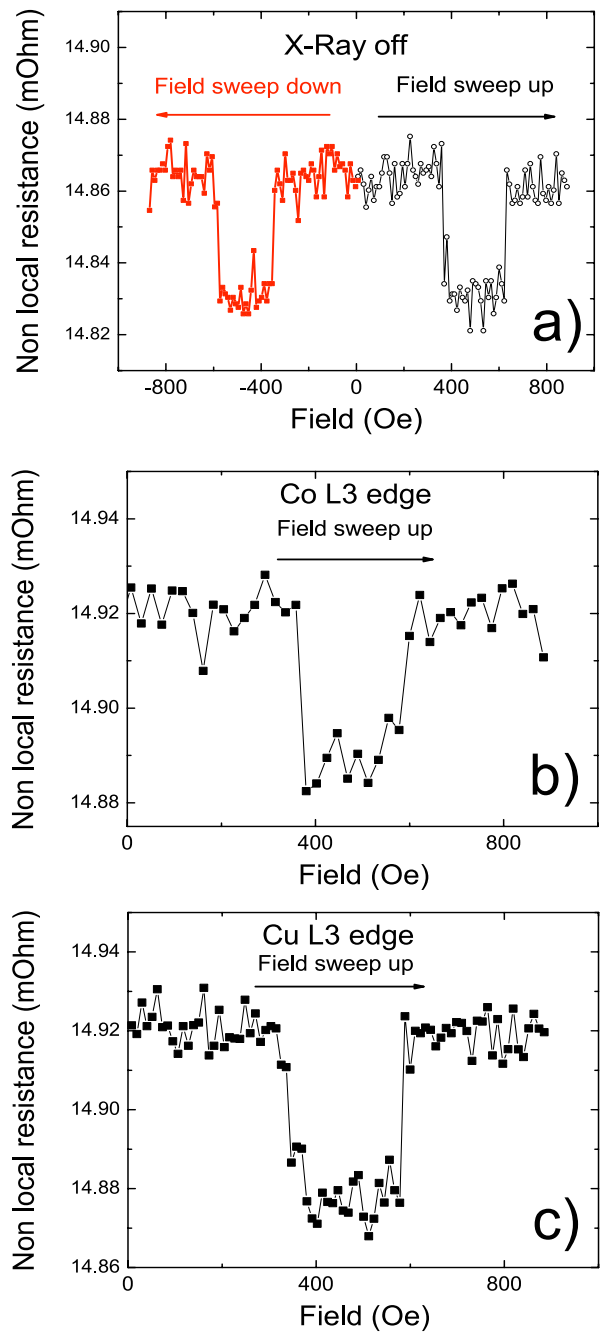}
 \caption{(Color online) Non local electrical measurements of
a Co/Cu lateral spin valve with a 200-nm separation.  The change in
the non-local resistance indicates a spin-accumulation in Cu and is
due to the parallel \textit{vs.} antiparallel orientation of the
magnetization in the two ferromagnetic contacts.  The measurements
were performed (a) without x-ray illumination, and while
illuminating at the (b) Co L$_3$, and (c) Cu $L_3$ absorption edges,
respectively.} \label{electric}
\end{figure}

Non-local resistance measurements in lateral spin valves have been a
powerful tool to investigate the spin injection from ferromagnetic
into non-magnetic materials
\cite{Johnson-PRL1985,Jedema-Nature2001,George-PRB2003,Kimura-APL2004,Hoffmann-APL2004,Crooker-Science2005,Hamrle-PRB2005,Ji-APL2006,Urech-NL2006,vanStaa-PRB2008,Casanova-PRB2009}.
A scanning electron microscope (SEM) image of a typical Co/Cu
lateral spin valve with 200-nm separation between ferromagnetic
contacts is shown in Fig.~\ref{SEM}.  The same figure shows the
schematic non-local transport measurement geometry, where a spin
polarized electrical current is injected from the Co into the Cu and
drained towards one end of the Cu wire, while the resultant spin
accumulation in the vicinity of the injection contact is measured
with a second Co contact, towards the opposite end, outside of the
direct path of the electrical current.  Measuring the voltage as a
function of the relative magnetization orientation of the two
ferromagnetic contacts gives rise to a voltage contrast that is
directly proportional to the spin accumulation
\cite{Johnson-PRL1985}.  This is shown in Fig.~\ref{electric}(a),
where the non-local resistance (the voltage normalized by the
injection current of 0.5~mA) is plotted as a function of increasing
and decreasing magnetic field. The different aspect ratios of the
contacts (350- and 200-nm width, respectively; as seen in
Fig.~\ref{SEM}) give rise to different switching fields and the
resultant non-local resistance contrast of $\approx 30~\mu\Omega$ is
consistent with earlier measurements on similar samples with a spin
diffusion length of 110~nm at room temperature
\cite{Ji-APL2006,Hoffmann-IOP2007}.

Apriori it is unclear whether the illumination with polarized x-rays
has any influence on the spin accumulation generated by electrical
injection.  For example, it is conceivable that the excitation with
circular polarized x-rays may either generate additional spin
accumulation (similar to optical spin injection in
semiconductors\cite{Awschalom-PRL98,Lampel-PRL68}) or result in an
enhanced spin relaxation. There is also the possibility that the
photoelectrons generated in the x-ray absorption process disturb the
balance in the spin accumulation process. To clarify this issue, we
performed non-local resistance measurements while circular polarized
x-rays with energies corresponding to the  Co L$_3$ and Cu L$_3$
absorption edges illuminated the sample.  As can be seen in
Figs.~\ref{electric}(b) and (c) the x-ray illumination resulted in a
small change of the base-signal (possibly due to generation of
photoelectrons), but the non-local resistance contrast upon
switching the magnetizations of the injection and detection contact
remained identical to the one observed without x-ray illumination.
This suggests that the impact of the impinging x-rays on the spin
accumulation in the Cu wire is negligible. However, one has to
realize that the electrical measurements are performed continuously
with a 10-s integration period, while the x-ray illumination is
pulsed with 65-ps x-ray pulse length and 2-ns pulse repetition.
Since the spin relaxation time in Cu is about 1~ps
\cite{Hoffmann-IOP2007}, any response to the x-ray illumination
should therefore be adiabatic and the electric measurements
effectively average over the illuminated (3.25\%) and dark (96.75\%)
periods.  Therefore, subtle changes of the electrical spin
accumulation signal due to the x-ray illumination, if any, would be
unresolved.

\begin{figure}
 \includegraphics[width=8.6cm]{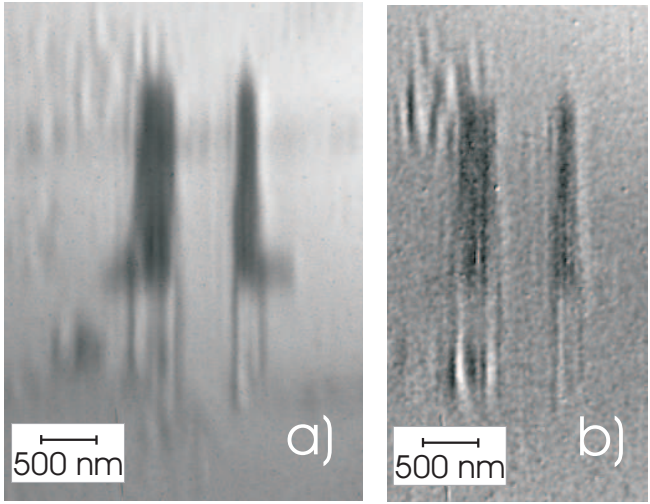}
 \caption{(a) TXM image of a Co/Cu lateral spin valve with a 500-nm separation taken at the Co L$_3$ absorption edge in remanence after saturation in a positive field.  The ferromagnetic Co electrodes are well defined in contrast to the barely visible Cu wires. (b) XMCD image of the same lateral Co/Cu spin valve. This image was obtained by subtracting images taken in remanence after positive and negative saturation, respectively. The dark XMCD contrast of $\sim$5\% indicates that the magnetization switched in both Co electrodes.} %
\label{Coedge}%
\end{figure}

For the imaging of spin accumulation in Cu we chose a Co/Cu lateral
spin valve with a 500-nm separation between injector and detector.
The larger distance between the contacts will result in less
structural interference of the second Co electrode with any
potential spin accumulation signal.  However, due to larger contact
separation we were not able to directly detect the spin accumulation
electrically with the sample mounted into the transmission X-ray
microscope. Nevertheless, we note that earlier electrical
measurements of samples with similar contact separation in a quieter
environment detected the spin accumulation in Cu \cite{Ji-APL2006}.
Furthermore, the electrical properties of the injector electrode
(resistance) were comparable to the 200-nm separation sample
discussed above. Figure~\ref{Coedge}(a) shows the TXM image taken at
the Co L$_3$ absorption edge in remanence after both electrodes were
saturated in a positive field.  Both ferromagnetic Co contacts are
well defined, while the contrast for the Cu wire is weak.  By
observing the XMCD contrast change in remanence we verified that a
magnetic field of $\pm840$~Oe reliably switches the magnetization of
both of the Co contacts.  This is shown in Fig.~\ref{Coedge}(b),
where images after positive and negative saturation were subtracted
from each other. Note that the dark magnetic contrast is homogeneous
in both electrodes, indicating that even in remanence the
magnetizations remain well defined in both electrodes, which is
consistent with the electrical measurements.

\begin{figure}%
\includegraphics[width=8.6cm]{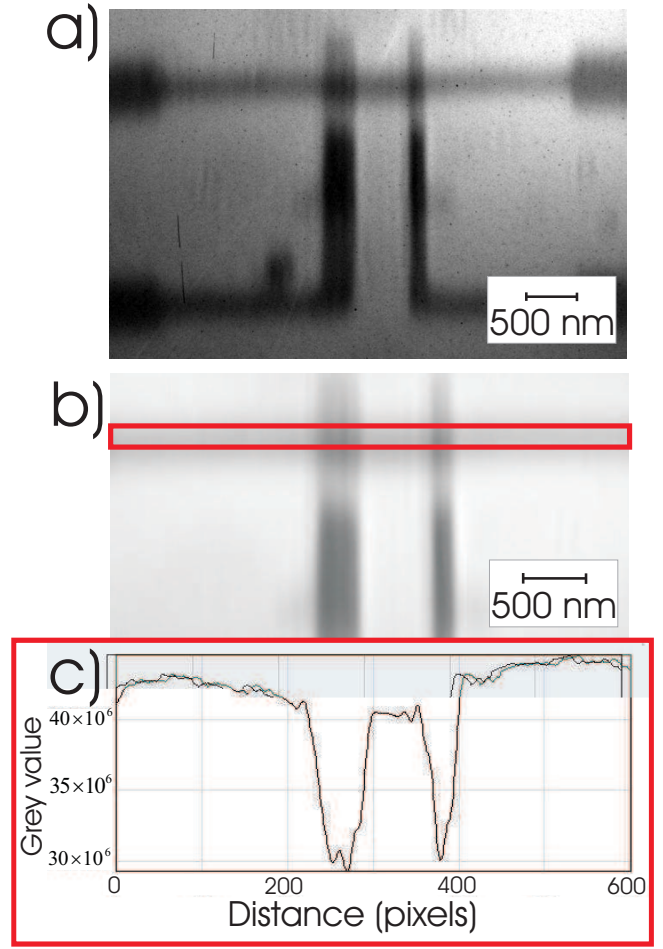}
 \caption{(Color online) (a) Single TXM image of a Co/Cu lateral spin valve with 500-nm separation taken at the Cu L$_3$ absorption edge in remanence after saturation at positive field.  There is a clear contrast for the Cu wires.  (b) Sum of 400 images taken at the Cu L$_3$ absorption edge.  The area of interest around the injection contact is magnified.  (c) X-ray intensity profile of the summed images along the Cu wire as indicated by the rectangular box in (b).  The additional attenuation of the Co electrodes is clearly visible.}
\label{Cuedge}%
\end{figure}

\begin{figure}%
\includegraphics[width=8.6cm]{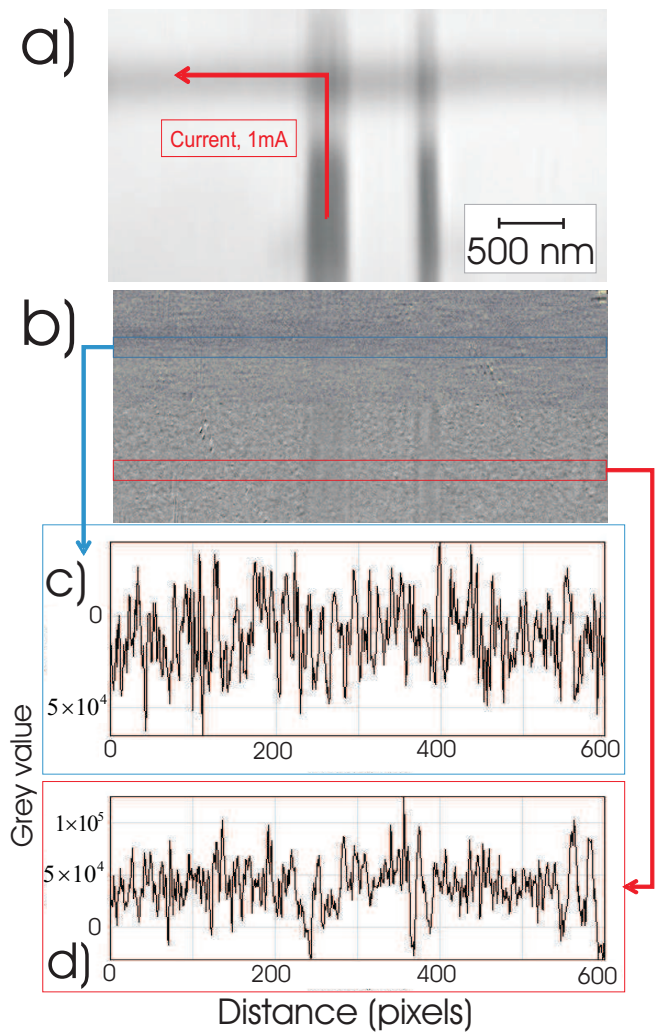}
\caption{(Color online) (a) Sum of 400 TXM images taken at the Cu L$_3$ absorption edge showing the location of the Co electrodes.  The electrical current path is also indicated.  (b)  XMCD image of the same region obtained by subtracting summed images after positive and negative saturation. No clear magnetic contrast is observed.  (c)  Intensity profile of the XMCD signal integrated over the width of the Cu-wire.  (d) Intensity profile of the XMCD signal integrated over the same width as in (c), but outside the region of the Cu wire.  Any intensity variations in (c) are comparable to the noise in (d).} %
\label{CuXMCD}%
\end{figure}

For investigating the spin accumulation in the Cu wire we performed
MTXM imaging at the Cu L$_3$ absorption edge.  During the image
recording a spin accumulation was continuously generated via
electrical injection with a {\em dc} 1-mA current through the wider
(left in Figs.~\ref{Coedge}--\ref{CuXMCD}) Co electrode.
Figure~\ref{Cuedge}(a) shows an individual TXM image.  In contrast
to the images at the Co L$_3$ absorption edge [see
Fig.~\ref{Coedge}(a)] the Cu wire and contacts are well defined by
the dark contrast.  In order to obtain maximum sensitivity for a
potentially small XMCD contrast of the spin accumulation, we
collected 400 images for each remanence with positive and negative
magnetization ($\sim$48~hours total measurement time).  In order to
compensate for a slow drift of the microscope optics, all individual
images were aligned using their structural contrast before being
summed up [see Fig.~\ref{Cuedge}(b)]. Figure~\ref{Cuedge}(c) shows a
line-profile of the x-ray intensity integrated across the width of
the Cu wire.  The positions of the Co electrodes are visible due to
their additional attenuation of the x-ray intensity.

In order to obtain an image with XMCD contrast we take the
difference of the individual sums of 400 images for positive and
negative saturation [see Fig.~\ref{CuXMCD}(b)].  The XMCD contrast
image does not show any distinct contrast within the area of the
injection contact [see Fig.~\ref{CuXMCD}(a) for the corresponding
structural contrast]. In order to analyze this result further we
show line-profiles of the XMCD signal integrated over the width of
the Cu wire [Fig.~\ref{CuXMCD}(c)] and in an adjacent region
[Fig.~\ref{CuXMCD}(d)] for comparison.  There is no signature of an
XMCD signal from the spin accumulation in the Cu wire exceeding the
noise level of data from the region without the Cu wire.

\begin{figure}%
\includegraphics[width=8.6cm]{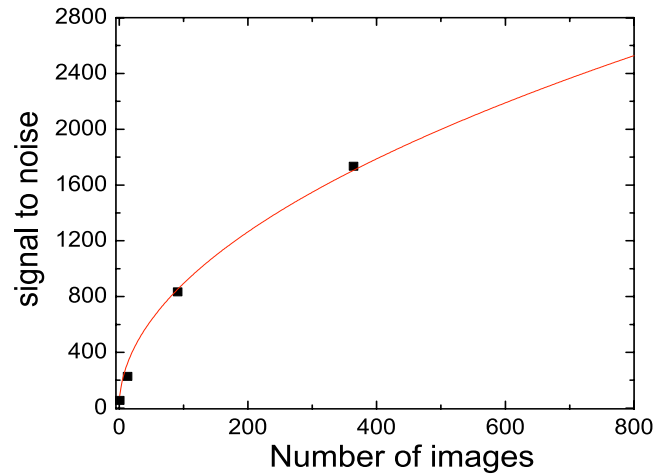}
\caption{Signal-to-noise ratio for each picture pixel plotted as a function of number $N$ of summed images. $\sqrt{N}$ fit is shown with solid line.} %
\label{noise}%
\end{figure}

One open question is if our signal-to-noise ratio is still limited
by the statistical noise.  We analyzed the signal-to-noise as a
function of the number $N$ of cumulated images by calculating the
ratio of the average intensity of the summed images over the noise
level in the subtracted XMCD images, see Fig.~\ref{noise}.  By
taking 400 images we were able to improve sensitivity of the
measurement in Cu by $\sim$100 times compared to the sensitivity of
a single image measurement in ferromagnetic Co. If the
signal-to-noise is dominated by statistical errors, it is expected
that that the signal-to-noise scales as $\sqrt{N}$.  As can be seen
in Fig.~\ref{noise} the square root fit matches the data well,
indicating that even after accumulating 400 images the
signal-to-noise is still dominated by counting statistics.
Furthermore the signal-to-noise is $\sim$1700 per pixel, meaning
that we are sensitive to a XMCD contrast of 0.06\% /pixel.  Any spin
accumulation should be present over an area given by the spin
diffusion length ($\approx 100$~nm) and thus should cover at least
15 pixels.  Therefore, any XMCD contrast from a spin accumulation in
Cu is $<$0.01\%. A further reduction of the signal-to-noise is not
practical. With the present experimental setup $\sim$48 h of data
accumulation were required with a current that is close to the
limiting current density for our devices.  Even with a 1~mA current
the lifetime of the lateral spin valve devices is limited by
electromigration and seldom exceed 2 days of a continuously applied
electrical current. Thus, presently, imaging the spin accumulation
with XMCD-based microscopic techniques does not seem to be a viable
option.

\begin{figure}%
\includegraphics[width=8.6cm]{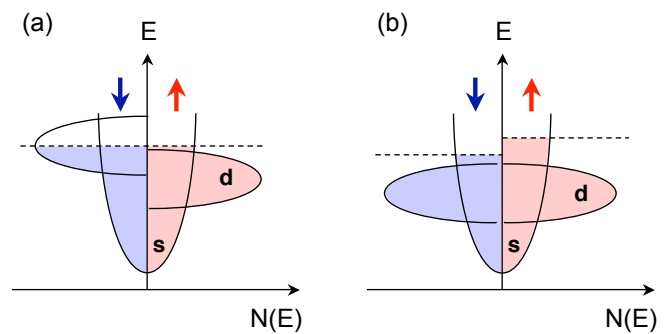}
\caption{(Color online) (a) Schematic band structure of a typical transition metal ferromagnet.  The XMCD contrast results from the spin-dependent density of states at the Fermi level and the concomitant different absorption as a function of photon polarization.  (b) Schematic band structure for spin accumulation in Cu.  The spin accumulation results in a spin-dependent splitting of the chemical potential, with nearly identical density of states.} %
\label{band}%
\end{figure}

Furthermore, the question remains whether spin accumulation should
indeed result in significant XMCD contrast.  The large XMCD contrast
for transition metal ferromagnets stems from the large exchange
splitting of the d-bands, which in turn gives rise to very different
spin-dependent density of states at the Fermi level [see
Fig.~\ref{band}(a)] and concomitantly a different magnitude of
absorption for different photon polarization states.  However, the
situation for spin accumulation is different, since the actual
band-structure is spin-independent and the only difference is a
splitting of the chemical potentials, as is schematically shown in
Fig.~\ref{band}(b).  Unless the density of states has a strong
energy dependence at the Fermi energy (which would be unexpected for
Cu) the absorption-rate for different photon polarizations will
therefore be similar, and the only difference is a slight
displacement of the absorption edge dependent on the photon
polarization.  From electrical measurements similar to the ones
presented in Fig.~\ref{electric} the measured voltage for the spin
accumulation is generally at most of order of $\mu$V (see
Ref.~\onlinecite{kimura-PRL2008}). Thus, even taking non-perfect
injection and detection efficiencies into account the splitting of
the chemical potentials can be expected to be well below 1~meV, and
thus significantly below the energy resolution ($\approx 1$~eV) of
our experiment.  Therefore, it remains doubtful that significant
XMCD contrast from a spin accumulation could be observed using the
approach discussed in this paper.

\section{Conclusions}
We fabricated Co/Cu lateral spin valves on free standing SiN
membranes for transmission soft x-ray microscopy imaging.  Non-local
resistivity measurements confirmed the presence of spin accumulation
in the Cu due to electrical injection from a Co electrode.  The
transport signature of the spin accumulation remained unchanged
under illumination with circularly polarized x-rays at both the Co
and Co L$_3$ absorption edges.  We obtained high resolution
transmission x-ray microscopy images at both the Co and Cu L$_3$
absorption edges.  However, no x-ray magnetic circular dichroism
signal from the spin accumulation in Cu was observed with a
signal-to-noise of 0.06\% per pixel.

\section{Acknowledgements}
We thank V. Yefremenko and V. Novosad for their help with the SiN
membrane fabrication. The CXRO and ALS staff is highly appreciated.
This work was supported by the Office of Basic Energy Sciences,
Materials Sciences and Engineering Division, of the U.S. Department
of Energy , under Contract Nos.\ DE-AC02-06CH11357 and
DE-AC02-05CH11231.

%\bibliography{refer}

\begin{thebibliography}{29}
\expandafter\ifx\csname
natexlab\endcsname\relax\def\natexlab#1{#1}\fi
\expandafter\ifx\csname bibnamefont\endcsname\relax
  \def\bibnamefont#1{#1}\fi
\expandafter\ifx\csname bibfnamefont\endcsname\relax
  \def\bibfnamefont#1{#1}\fi
\expandafter\ifx\csname citenamefont\endcsname\relax
  \def\citenamefont#1{#1}\fi
\expandafter\ifx\csname url\endcsname\relax
  \def\url#1{\texttt{#1}}\fi
\expandafter\ifx\csname urlprefix\endcsname\relax\def\urlprefix{URL
}\fi \providecommand{\bibinfo}[2]{#2}
\providecommand{\eprint}[2][]{\url{#2}}

\bibitem[{\citenamefont{Chappert and Kim}(2008)}]{Chappert-NP2008}
\bibinfo{author}{\bibfnamefont{C.}~\bibnamefont{Chappert}} \bibnamefont{and}
  \bibinfo{author}{\bibfnamefont{J.-V.} \bibnamefont{Kim}},
  \bibinfo{journal}{Nat. Phys.} \textbf{\bibinfo{volume}{4}},
  \bibinfo{pages}{837} (\bibinfo{year}{2008}).

\bibitem[{\citenamefont{Hoffmann}(2007)}]{Hoffmann-PSSC2007}
\bibinfo{author}{\bibfnamefont{A.}~\bibnamefont{Hoffmann}},
  \bibinfo{journal}{Phys. Stat. Sol. (c)} \textbf{\bibinfo{volume}{4}},
  \bibinfo{pages}{4236} (\bibinfo{year}{2007}).

\bibitem[{\citenamefont{Johnson and Silsbee}(1985)}]{Johnson-PRL1985}
\bibinfo{author}{\bibfnamefont{M.}~\bibnamefont{Johnson}} \bibnamefont{and}
  \bibinfo{author}{\bibfnamefont{R.~H.} \bibnamefont{Silsbee}},
  \bibinfo{journal}{Phys. Rev. Lett.} \textbf{\bibinfo{volume}{55}},
  \bibinfo{pages}{1790} (\bibinfo{year}{1985}).

\bibitem[{\citenamefont{Kimura et~al.}(2007)\citenamefont{Kimura, Otani, Sato,
  Takahashi, and Maekawa}}]{Kimura-PRL2007}
\bibinfo{author}{\bibfnamefont{T.}~\bibnamefont{Kimura}},
  \bibinfo{author}{\bibfnamefont{Y.}~\bibnamefont{Otani}},
  \bibinfo{author}{\bibfnamefont{T.}~\bibnamefont{Sato}},
  \bibinfo{author}{\bibfnamefont{S.}~\bibnamefont{Takahashi}},
  \bibnamefont{and} \bibinfo{author}{\bibfnamefont{S.}~\bibnamefont{Maekawa}},
  \bibinfo{journal}{Phys. Rev. Lett.} \textbf{\bibinfo{volume}{98}},
  \bibinfo{eid}{156601} (\bibinfo{year}{2007}).

\bibitem[{\citenamefont{Heinrich et~al.}(2003)\citenamefont{Heinrich,
  Tserkovnyak, Woltersdorf, Brataas, Urban, and Bauer}}]{Heinrich-PRL2003}
\bibinfo{author}{\bibfnamefont{B.}~\bibnamefont{Heinrich}},
  \bibinfo{author}{\bibfnamefont{Y.}~\bibnamefont{Tserkovnyak}},
  \bibinfo{author}{\bibfnamefont{G.}~\bibnamefont{Woltersdorf}},
  \bibinfo{author}{\bibfnamefont{A.}~\bibnamefont{Brataas}},
  \bibinfo{author}{\bibfnamefont{R.}~\bibnamefont{Urban}}, \bibnamefont{and}
  \bibinfo{author}{\bibfnamefont{G.}~\bibnamefont{Bauer}},
  \bibinfo{journal}{PRL} \textbf{\bibinfo{volume}{90}}, \bibinfo{pages}{187601}
  (\bibinfo{year}{2003}).

\bibitem[{\citenamefont{Woltersdorf et~al.}(2007)\citenamefont{Woltersdorf,
  Mosendz, Heinrich, and Back}}]{Woltersdorf-PRL2007}
\bibinfo{author}{\bibfnamefont{G.}~\bibnamefont{Woltersdorf}},
  \bibinfo{author}{\bibfnamefont{O.}~\bibnamefont{Mosendz}},
  \bibinfo{author}{\bibfnamefont{B.}~\bibnamefont{Heinrich}}, \bibnamefont{and}
  \bibinfo{author}{\bibfnamefont{C.}~\bibnamefont{Back}},
  \bibinfo{journal}{Phys. Rev. Lett.} \textbf{\bibinfo{volume}{99}},
  \bibinfo{pages}{246603} (\bibinfo{year}{2007}).

\bibitem[{\citenamefont{Mosendz et~al.}(2009)\citenamefont{Mosendz,
  Woltersdorf, Kardasz, Heinrich, and Back}}]{Mosendz-PRB09}
\bibinfo{author}{\bibfnamefont{O.}~\bibnamefont{Mosendz}},
  \bibinfo{author}{\bibfnamefont{G.}~\bibnamefont{Woltersdorf}},
  \bibinfo{author}{\bibfnamefont{B.}~\bibnamefont{Kardasz}},
  \bibinfo{author}{\bibfnamefont{B.}~\bibnamefont{Heinrich}}, \bibnamefont{and}
  \bibinfo{author}{\bibfnamefont{C.}~\bibnamefont{Back}},
  \bibinfo{journal}{Phys. Rev. B} \textbf{\bibinfo{volume}{79}},
  \bibinfo{pages}{224412} (\bibinfo{year}{2009}).

\bibitem[{\citenamefont{Kikkawa and Awschalom}(1998)}]{Awschalom-PRL98}
\bibinfo{author}{\bibfnamefont{J.}~\bibnamefont{Kikkawa}} \bibnamefont{and}
  \bibinfo{author}{\bibfnamefont{D.}~\bibnamefont{Awschalom}},
  \bibinfo{journal}{Phys. Rev. Lett.} \textbf{\bibinfo{volume}{80}},
  \bibinfo{pages}{4313} (\bibinfo{year}{1998}).

\bibitem[{\citenamefont{Kato et~al.}(2004)\citenamefont{Kato, Myers, Gossard,
  and Awschalom}}]{Awschalom-SC04}
\bibinfo{author}{\bibfnamefont{Y.}~\bibnamefont{Kato}},
  \bibinfo{author}{\bibfnamefont{R.}~\bibnamefont{Myers}},
  \bibinfo{author}{\bibfnamefont{A.}~\bibnamefont{Gossard}}, \bibnamefont{and}
  \bibinfo{author}{\bibfnamefont{D.}~\bibnamefont{Awschalom}},
  \bibinfo{journal}{Science} \textbf{\bibinfo{volume}{306}},
  \bibinfo{pages}{1910} (\bibinfo{year}{2004}).

\bibitem[{\citenamefont{Sih et~al.}(2005)\citenamefont{Sih, Myers, Kato, Lau,
  Gossard, and Awschalom}}]{Awschalom-NP05}
\bibinfo{author}{\bibfnamefont{V.}~\bibnamefont{Sih}},
  \bibinfo{author}{\bibfnamefont{R.}~\bibnamefont{Myers}},
  \bibinfo{author}{\bibfnamefont{Y.}~\bibnamefont{Kato}},
  \bibinfo{author}{\bibfnamefont{W.}~\bibnamefont{Lau}},
  \bibinfo{author}{\bibfnamefont{A.}~\bibnamefont{Gossard}}, \bibnamefont{and}
  \bibinfo{author}{\bibfnamefont{D.}~\bibnamefont{Awschalom}},
  \bibinfo{journal}{Nature Physis} \textbf{\bibinfo{volume}{1}},
  \bibinfo{pages}{31} (\bibinfo{year}{2005}).

\bibitem[{\citenamefont{Crooker et~al.}(2005)\citenamefont{Crooker, Furis, Lou,
  Adelmann, Smith, Palmstrom, and Crowell}}]{Crooker-Science2005}
\bibinfo{author}{\bibfnamefont{S.~A.} \bibnamefont{Crooker}},
  \bibinfo{author}{\bibfnamefont{M.}~\bibnamefont{Furis}},
  \bibinfo{author}{\bibfnamefont{X.}~\bibnamefont{Lou}},
  \bibinfo{author}{\bibfnamefont{C.}~\bibnamefont{Adelmann}},
  \bibinfo{author}{\bibfnamefont{D.~L.} \bibnamefont{Smith}},
  \bibinfo{author}{\bibfnamefont{C.~J.} \bibnamefont{Palmstrom}},
  \bibnamefont{and} \bibinfo{author}{\bibfnamefont{P.~A.}
  \bibnamefont{Crowell}}, \bibinfo{journal}{Science}
  \textbf{\bibinfo{volume}{309}}, \bibinfo{pages}{2191} (\bibinfo{year}{2005}).

\bibitem[{\citenamefont{Stephens et~al.}(2004)\citenamefont{Stephens,
  Berezovsky, McGuire, Sham, Gossard, and Awschalom}}]{Stephens-PRL2005}
\bibinfo{author}{\bibfnamefont{J.}~\bibnamefont{Stephens}},
  \bibinfo{author}{\bibfnamefont{J.}~\bibnamefont{Berezovsky}},
  \bibinfo{author}{\bibfnamefont{J.~P.} \bibnamefont{McGuire}},
  \bibinfo{author}{\bibfnamefont{L.~J.} \bibnamefont{Sham}},
  \bibinfo{author}{\bibfnamefont{A.~C.} \bibnamefont{Gossard}},
  \bibnamefont{and} \bibinfo{author}{\bibfnamefont{D.~D.}
  \bibnamefont{Awschalom}}, \bibinfo{journal}{Phys. Rev. Lett.}
  \textbf{\bibinfo{volume}{93}}, \bibinfo{pages}{097602}
  (\bibinfo{year}{2004}).

\bibitem[{\citenamefont{Bass and Pratt}(2007)}]{bass-review}
\bibinfo{author}{\bibfnamefont{J.}~\bibnamefont{Bass}} \bibnamefont{and}
  \bibinfo{author}{\bibfnamefont{W.}~\bibnamefont{Pratt}},
  \bibinfo{journal}{Journal of Physics: Condenced Matter}
  \textbf{\bibinfo{volume}{19}}, \bibinfo{pages}{183201}
  (\bibinfo{year}{2007}).

\bibitem[{\citenamefont{Urech et~al.}(2006)\citenamefont{Urech, Korenivski,
  Poli, and Haviland}}]{Urech-NL2006}
\bibinfo{author}{\bibfnamefont{M.}~\bibnamefont{Urech}},
  \bibinfo{author}{\bibfnamefont{V.}~\bibnamefont{Korenivski}},
  \bibinfo{author}{\bibfnamefont{N.}~\bibnamefont{Poli}}, \bibnamefont{and}
  \bibinfo{author}{\bibfnamefont{D.~B.} \bibnamefont{Haviland}},
  \bibinfo{journal}{Nano Lett.} \textbf{\bibinfo{volume}{6}},
  \bibinfo{pages}{871} (\bibinfo{year}{2006}).

\bibitem[{\citenamefont{Hamrle et~al.}(2005)\citenamefont{Hamrle, Kimura,
  Otani, Tsukagoshi, and Aoyagi}}]{Hamrle-PRB2005}
\bibinfo{author}{\bibfnamefont{J.}~\bibnamefont{Hamrle}},
  \bibinfo{author}{\bibfnamefont{T.}~\bibnamefont{Kimura}},
  \bibinfo{author}{\bibfnamefont{Y.}~\bibnamefont{Otani}},
  \bibinfo{author}{\bibfnamefont{K.}~\bibnamefont{Tsukagoshi}},
  \bibnamefont{and} \bibinfo{author}{\bibfnamefont{Y.}~\bibnamefont{Aoyagi}},
  \bibinfo{journal}{Phys. Rev. B} \textbf{\bibinfo{volume}{71}},
  \bibinfo{eid}{094402} (\bibinfo{year}{2005}).

\bibitem[{\citenamefont{Fischer}(2008)}]{Fischer-IEEE08}
\bibinfo{author}{\bibfnamefont{P.}~\bibnamefont{Fischer}},
  \bibinfo{journal}{IEEE Trans. Magn.} \textbf{\bibinfo{volume}{44}},
  \bibinfo{pages}{1900} (\bibinfo{year}{2008}).

\bibitem[{\citenamefont{Chao et~al.}(2005)\citenamefont{Chao, Harteneck,
  Liddle, Anderson, and Attwood}}]{Chao-Nature2005}
\bibinfo{author}{\bibfnamefont{W.}~\bibnamefont{Chao}},
  \bibinfo{author}{\bibfnamefont{B.~D.} \bibnamefont{Harteneck}},
  \bibinfo{author}{\bibfnamefont{J.~A.} \bibnamefont{Liddle}},
  \bibinfo{author}{\bibfnamefont{E.~H.} \bibnamefont{Anderson}},
  \bibnamefont{and} \bibinfo{author}{\bibfnamefont{D.~T.}
  \bibnamefont{Attwood}}, \bibinfo{journal}{Nature}
  \textbf{\bibinfo{volume}{435}}, \bibinfo{pages}{1213} (\bibinfo{year}{2005}).

\bibitem[{\citenamefont{Fischer et~al.}(2006)\citenamefont{Fischer, Kim, Chao,
  Liddle, Anderson, and Attwood}}]{Fischer-MT2006}
\bibinfo{author}{\bibfnamefont{P.}~\bibnamefont{Fischer}},
  \bibinfo{author}{\bibfnamefont{D.-H.} \bibnamefont{Kim}},
  \bibinfo{author}{\bibfnamefont{W.}~\bibnamefont{Chao}},
  \bibinfo{author}{\bibfnamefont{J.~A.} \bibnamefont{Liddle}},
  \bibinfo{author}{\bibfnamefont{E.~H.} \bibnamefont{Anderson}},
  \bibnamefont{and} \bibinfo{author}{\bibfnamefont{D.~T.}
  \bibnamefont{Attwood}}, \bibinfo{journal}{Mater. Today}
  \textbf{\bibinfo{volume}{9}}, \bibinfo{pages}{26 } (\bibinfo{year}{2006}).

\bibitem[{\citenamefont{Cord et~al.}(2006)\citenamefont{Cord, Dames, Berggren,
  and Aumentado}}]{Cord-JVST2006}
\bibinfo{author}{\bibfnamefont{B.}~\bibnamefont{Cord}},
  \bibinfo{author}{\bibfnamefont{C.}~\bibnamefont{Dames}},
  \bibinfo{author}{\bibfnamefont{K.~K.} \bibnamefont{Berggren}},
  \bibnamefont{and}
  \bibinfo{author}{\bibfnamefont{J.}~\bibnamefont{Aumentado}},
  \bibinfo{journal}{J. Vac. Sci. Technol. B} \textbf{\bibinfo{volume}{24}},
  \bibinfo{pages}{3139} (\bibinfo{year}{2006}).

\bibitem[{\citenamefont{Jedema et~al.}(2001)\citenamefont{Jedema, Filip, and
  van Wees}}]{Jedema-Nature2001}
\bibinfo{author}{\bibfnamefont{F.~J.} \bibnamefont{Jedema}},
  \bibinfo{author}{\bibfnamefont{A.~T.} \bibnamefont{Filip}}, \bibnamefont{and}
  \bibinfo{author}{\bibfnamefont{B.~J.} \bibnamefont{van Wees}},
  \bibinfo{journal}{Nature} \textbf{\bibinfo{volume}{410}},
  \bibinfo{pages}{345} (\bibinfo{year}{2001}).

\bibitem[{\citenamefont{George et~al.}(2003)\citenamefont{George, Fert, and
  Faini}}]{George-PRB2003}
\bibinfo{author}{\bibfnamefont{J.-M.} \bibnamefont{George}},
  \bibinfo{author}{\bibfnamefont{A.}~\bibnamefont{Fert}}, \bibnamefont{and}
  \bibinfo{author}{\bibfnamefont{G.}~\bibnamefont{Faini}},
  \bibinfo{journal}{Phys. Rev. B} \textbf{\bibinfo{volume}{67}},
  \bibinfo{pages}{012410} (\bibinfo{year}{2003}).

\bibitem[{\citenamefont{Kimura et~al.}(2004)\citenamefont{Kimura, Hamrle,
  Otani, Tsukagoshi, and Aoyagi}}]{Kimura-APL2004}
\bibinfo{author}{\bibfnamefont{T.}~\bibnamefont{Kimura}},
  \bibinfo{author}{\bibfnamefont{J.}~\bibnamefont{Hamrle}},
  \bibinfo{author}{\bibfnamefont{Y.}~\bibnamefont{Otani}},
  \bibinfo{author}{\bibfnamefont{K.}~\bibnamefont{Tsukagoshi}},
  \bibnamefont{and} \bibinfo{author}{\bibfnamefont{Y.}~\bibnamefont{Aoyagi}},
  \bibinfo{journal}{Appl. Phys. Lett.} \textbf{\bibinfo{volume}{85}},
  \bibinfo{pages}{3501} (\bibinfo{year}{2004}).

\bibitem[{\citenamefont{Ji et~al.}(2004)\citenamefont{Ji, Hoffmann, Jiang, and
  Bader}}]{Hoffmann-APL2004}
\bibinfo{author}{\bibfnamefont{Y.}~\bibnamefont{Ji}},
  \bibinfo{author}{\bibfnamefont{A.}~\bibnamefont{Hoffmann}},
  \bibinfo{author}{\bibfnamefont{J.~S.} \bibnamefont{Jiang}}, \bibnamefont{and}
  \bibinfo{author}{\bibfnamefont{S.~D.} \bibnamefont{Bader}},
  \bibinfo{journal}{Appl. Phys. Lett.} \textbf{\bibinfo{volume}{85}},
  \bibinfo{pages}{6218} (\bibinfo{year}{2004}).

\bibitem[{\citenamefont{Ji et~al.}(2006)\citenamefont{Ji, Hoffmann, Pearson,
  and Bader}}]{Ji-APL2006}
\bibinfo{author}{\bibfnamefont{Y.}~\bibnamefont{Ji}},
  \bibinfo{author}{\bibfnamefont{A.}~\bibnamefont{Hoffmann}},
  \bibinfo{author}{\bibfnamefont{J.~E.} \bibnamefont{Pearson}},
  \bibnamefont{and} \bibinfo{author}{\bibfnamefont{S.~D.} \bibnamefont{Bader}},
  \bibinfo{journal}{Applied Physics Letters} \textbf{\bibinfo{volume}{88}},
  \bibinfo{eid}{052509} (\bibinfo{year}{2006}).

\bibitem[{\citenamefont{van Staa et~al.}(2008)\citenamefont{van Staa,
  Wulfhorst, Vogel, Merkt, and Meier}}]{vanStaa-PRB2008}
\bibinfo{author}{\bibfnamefont{A.}~\bibnamefont{van Staa}},
  \bibinfo{author}{\bibfnamefont{J.}~\bibnamefont{Wulfhorst}},
  \bibinfo{author}{\bibfnamefont{A.}~\bibnamefont{Vogel}},
  \bibinfo{author}{\bibfnamefont{U.}~\bibnamefont{Merkt}}, \bibnamefont{and}
  \bibinfo{author}{\bibfnamefont{G.}~\bibnamefont{Meier}},
  \bibinfo{journal}{Phys. Rev. B} \textbf{\bibinfo{volume}{77}},
  \bibinfo{eid}{214416} (\bibinfo{year}{2008}).

\bibitem[{\citenamefont{Casanova et~al.}(2009)\citenamefont{Casanova, Sharoni,
  Erekhinsky, and Schuller}}]{Casanova-PRB2009}
\bibinfo{author}{\bibfnamefont{F.}~\bibnamefont{Casanova}},
  \bibinfo{author}{\bibfnamefont{A.}~\bibnamefont{Sharoni}},
  \bibinfo{author}{\bibfnamefont{M.}~\bibnamefont{Erekhinsky}},
  \bibnamefont{and} \bibinfo{author}{\bibfnamefont{I.~K.}
  \bibnamefont{Schuller}}, \bibinfo{journal}{Phys. Rev. B}
  \textbf{\bibinfo{volume}{79}}, \bibinfo{eid}{184415} (\bibinfo{year}{2009}).

\bibitem[{\citenamefont{Ji et~al.}(2007)\citenamefont{Ji, Hoffmann, Jiang,
  Pearson, and Bader}}]{Hoffmann-IOP2007}
\bibinfo{author}{\bibfnamefont{Y.}~\bibnamefont{Ji}},
  \bibinfo{author}{\bibfnamefont{A.}~\bibnamefont{Hoffmann}},
  \bibinfo{author}{\bibfnamefont{J.~S.} \bibnamefont{Jiang}},
  \bibinfo{author}{\bibfnamefont{J.~E.} \bibnamefont{Pearson}},
  \bibnamefont{and} \bibinfo{author}{\bibfnamefont{S.~D.} \bibnamefont{Bader}},
  \bibinfo{journal}{J. Phys. D: Appl. Phys.} \textbf{\bibinfo{volume}{40}},
  \bibinfo{pages}{1280} (\bibinfo{year}{2007}).

\bibitem[{\citenamefont{Lampel}(1968)}]{Lampel-PRL68}
\bibinfo{author}{\bibfnamefont{G.}~\bibnamefont{Lampel}},
  \bibinfo{journal}{Phys. Rev. Lett.} \textbf{\bibinfo{volume}{20}},
  \bibinfo{pages}{491} (\bibinfo{year}{1968}).

\bibitem[{\citenamefont{Kimura et~al.}(2008)\citenamefont{Kimura, Sato, and
  Otani}}]{kimura-PRL2008}
\bibinfo{author}{\bibfnamefont{T.}~\bibnamefont{Kimura}},
  \bibinfo{author}{\bibfnamefont{T.}~\bibnamefont{Sato}}, \bibnamefont{and}
  \bibinfo{author}{\bibfnamefont{Y.}~\bibnamefont{Otani}},
  \bibinfo{journal}{Phys. Rev. Lett.} \textbf{\bibinfo{volume}{100}},
  \bibinfo{pages}{066602} (\bibinfo{year}{2008}).

\end{thebibliography}

\end{document}